\documentclass[twocolumn,secnumarabic,amssymb, nobibnotes, aps, prx,10pt, superscriptaddress, nobalancelastpage,longbibliography]{revtex4-2}

\newcommand{\nuc}[2]{\hbox{$^{#1}$#2}}
 \usepackage{graphicx}
\usepackage{bm}
\usepackage{amsmath}
\usepackage{braket}
\usepackage{epsfig}
\usepackage{tensor}
\usepackage{CJKutf8}
\usepackage[version=4]{mhchem}
\usepackage{titlesec}
\usepackage{ifthen}
\usepackage{sidecap}
\usepackage{listings}
\usepackage{array}
\usepackage{floatrow}
\usepackage{enumitem}
\usepackage[para,online,flushleft]{threeparttablex}
\usepackage{orcidlink}
\usepackage{float}
\floatstyle{plaintop}
\restylefloat{table}

\usepackage{hyperref}
\hypersetup{breaklinks=true,colorlinks=true,linkcolor=blue,citecolor=blue,filecolor=magenta,urlcolor=cyan}

\usepackage[all]{hypcap}

\usepackage{xcolor}
\definecolor{pastelgray}{rgb}{0.81, 0.81, 0.77}
\definecolor{beaublue}{rgb}{0.9, 0.9, 0.93}

\makeatletter
\def\@bibdataout@aps{%
\immediate\write\@bibdataout{%
@CONTROL{%
apsrev41Control%
\longbibliography@sw{%
    ,author="08",editor="1",pages="1",title="0",year="1"%
    }{%
    ,author="08",editor="1",pages="1",title="",year="1"%
    }%
  }%
}%
\if@filesw \immediate \write \@auxout {\string \citation {apsrev41Control}}\fi
}
\makeatother

\newcolumntype{Y}{>{\centering\arraybackslash}X}

\begin{document}
\begin{CJK*}{UTF8}{gbsn}

\title{Extraction of ground-state nuclear deformations from ultra-relativistic heavy-ion collisions: Nuclear structure physics context}

\author{J. Dobaczewski\orcidlink{0000-0002-4158-3770}}
\email[]{jacek.dobaczewski@york.ac.uk}
\affiliation{School of Physics, Engineering and Technology University of York, Heslington, York YO10 5DD, United Kingdom}
\affiliation{Institute of Theoretical Physics, Faculty of Physics, University of Warsaw, ul. Pasteura 5, PL-02-093 Warsaw, Poland}

\author{A. Gade\,\orcidlink{0000-0001-8825-0976}}
\email[]{gade@frib.msu.edu}
\affiliation{Facility for Rare Isotope Beams, Michigan State University, East Lansing, Michigan 48824, USA}
\affiliation{Department of Physics and Astronomy, Michigan State University, East Lansing, Michigan 48824, USA}

\author{K. Godbey\,\orcidlink{0000-0003-0622-3646}}
\email[]{godbey@frib.msu.edu}
\affiliation{Facility for Rare Isotope Beams, Michigan State University, East Lansing, Michigan 48824, USA}

\author{R.V.F. Janssens\orcidlink{0000-0001-7095-1715}}
\email[]{rvfj@email.unc.edu}
\affiliation{Department of Physics and Astronomy, University of North Carolina at Chapel Hill, Chapel Hill, NC 27559-3255, USA and Triangle Universities Nuclear Laboratory, Duke University, Durham, NC 27708-0308, USA}

\author{W. Nazarewicz\,\orcidlink{0000-0002-8084-7425}}
\email[]{witek@frib.msu.edu}
\affiliation{Facility for Rare Isotope Beams, Michigan State University, East Lansing, Michigan 48824, USA}
\affiliation{Department of Physics and Astronomy, Michigan State University, East Lansing, Michigan 48824, USA}

\begin{abstract}
The collective-flow-assisted nuclear shape-imaging method in ultra-relativistic heavy-ion collisions has recently been used to characterize nuclear collective states. In this paper, we assess the foundations of the shape-imaging technique employed in these studies. We argue that some current UHIC nuclear imaging techniques neglect fundamental aspects of spontaneous symmetry-breaking and symmetry-restoration in colliding ions and incorrectly infer one-body multipole moments from studies of nucleonic correlations. Therefore, the impact of this approach on nuclear structure research has been overstated.
Conversely, efforts to incorporate existing knowledge on nuclear shapes into analysis pipelines can be beneficial for benchmarking tools and calibrating models used to extract information from ultra-relativistic heavy-ion  experiments.
\end{abstract}

\date{\today}

\maketitle
\end{CJK*}

\section{Introduction}\label{Introduction}

The shapes of nuclei  colliding in ultra-relativistic heavy-ion collisions (UHIC) influence the geometry of the quark-gluon plasma (QGP) created, which, in turn, affects the momentum distribution of the particles produced \cite{Alver2008,Shuryak2017,Busza2018}. The premise of the UHIC studies is rooted in the fact that
the resulting anisotropic (hydrodynamic) expansion of the QGP  \cite{Heinz2024} converts the initial spatial asymmetries into momentum anisotropies of the measured particles.
A typical method of analyzing this flow consists in characterizing the azimuthal anisotropy by the Fourier coefficients $\nu_n$ of the flow \cite{Adamczyk2015}; the elliptic flow, $\nu_2$, is sensitive to the ellipticity of the QGP, while the triangular one, $\nu_3$, is sensitive to the triangularity, and so on.

By relating the flow anisotropies to the geometry of the QGP initial state, a  shape-imaging method based on  UHIC experiments   at the Relativistic Heavy Ion Collider (RHIC) and the Large Hadron Collider (LHC) has been suggested as a tool to study  the geometrical shapes of the colliding
nuclei~\cite{Giacalone2020,Giacalone2020a,Giacalone2021,Giacalone2021a,Jia2022,(Jia22),Zhang2022,Bally2022,Bally2023,Ryssens2023,Abdulhamid2024,Jia2024,STAR2025,Giacalone2025,Giacalone2025b,Li2025,Fortier2025,Liu2025}.
In the present study, we examine the assumptions underlying the proposed shape-imaging technique in greater detail. In particular, we critically assess the claim in Ref.~\cite{Jia2024} that this approach ``not only refines our
understanding of the initial stages of the collisions but also turns high-energy nuclear experiments into a precision tool for imaging nuclear structures.''

The paper is organized as follows.  Section~\ref{snapshots} discusses essential concepts of symmetry breaking and restoration in the context of the two-body densities and Sec.~\ref{fluctuations} covers the notion of ``shape fluctuations.'' In Sec.~\ref{deformations} we recall the vast body of available knowledge on nuclear shapes from low-energy measurements.
The questions related to the uncertainty quantification (UQ) of the shape-imaging method are raised in Sec.~\ref{UQ}. The conclusions of the paper are given in Sec.~\ref{Conclusions}.

\section{Symmetries: breaking and restoration \label{snapshots}}

Let us first examine the symmetry aspects of self-bound many-body systems such as atomic nuclei.
A fundamental theorem of quantum mechanics states that an eigenstate of a symmetry-conserving Hamiltonian must belong to a representation of the symmetry group. Therefore, the ground states of $^{238}$U and $^{20}$Ne, which a rotationally-invariant nuclear Hamiltonian governs, belong to the $J^\pi=0^+$ representation, i.e., they are invariant under spatial orientation and reflections. In other words, they are perfectly spherical states
in the beam's reference frame (or perfectly axial in the laboratory frame due to the huge relativistic contraction). Similarly, any ground or excited state with $J>1/2$ \cite{Bragg1948,Bohr1951} for which the laboratory quadrupole moment is non-zero is covariant under spatial orientation (its rotation mixes the magnetic substates in a specific way). We focus on the $J^\pi=0^+$ case for the following discussion, but note that all arguments apply to any of the others.

What is the shape, then, of the $^{238}$U nucleus in its 0$^+$ ground state and how can it be measured?
The answer to these questions is irrespective of whether we use UHIC, electron scattering, or Coulomb excitation.
A concept that addresses this properly is spontaneous symmetry breaking -- an essential notion ubiquitous across all the various domains of physics. As exposed by Anderson in his seminal article~\cite{(And72)}, it encompasses finite systems (such as molecules or nuclei), very large systems (like crystals), and infinite systems (such as quantum fields). It does not contradict the aforementioned fundamental theorem of quantum mechanics; on the contrary, it allows us to understand what it means to measure the shape of a quantum object.

A false friend in this understanding is an intuition that allows one to picture the deformed $^{238}$U nucleus as a rotating and vibrating classical object, with corresponding time scales inferred from the energies of the quantum rotational and vibrational states~\cite{Jia2022,Zhang2022,Abdulhamid2024,Jia2024,STAR2025}. This is a misleading picture -- the $0^+$ ground state of $^{238}$U studied in the experiments is a stationary eigenstate; it does not rotate or vibrate, it does not change with time, and, thus, no instantaneous snapshot of the shape can be taken, no matter the time scale of the experiment performed.

A proper quantum-mechanical intuition that helps to understand the shape of the $^{238}$U 0$^+$ ground state is a model of a spontaneously symmetry-broken deformed wave function. This is not the wave function that reaches the detector in the UHIC experiment. However, this deformed wave function can be used to build {\em an intuitive model} of the $^{238}$U 0$^+$ ground state by constructing a linear combination of wave functions differently oriented in space and having different shapes, that is, by restoring the symmetry~\cite{Sheikh21}. This model of the 0$^+$ ground state is still time-independent; it does not rotate or vibrate. Most importantly, one must remember that in this model state, the wave functions of different orientations and shapes are mixed, not their squares or density distributions, as shown in diagrams like the tip-tip or body-body configurations of Ref.~\cite{STAR2025}. Recently, this aspect has been further explored quantitatively in Ref.~\cite{Ke2025}.

As is the case for any other quantum-mechanical wave function, a {\em single} experiment, when properly designed, may project the wave function on a given value of its ``position" (its orientation and shape). However, a series of several identical experiments will give {\em different} values of positions. In particular, for the 0$^+$ $^{238}$U ground state, all the different possible orientations will come out with equal probabilities, and the information on the shape will be inaccessible. Indeed, it must be so, as the shape of the 0$^+$ state is spherical.

A similar discussion can be made about the ``bowling pin'' schematics
of Refs.~\cite{Giacalone2025,Giacalone2025b} symbolizing a pear-shaped nucleus, $^{20}$Ne. Indeed,  in the laboratory system, the $J^\pi=0^+$ ground-state wave function of  $^{20}$Ne is perfectly spherical and reflection-symmetric, i.e., it does not point in a specific direction.

So, how then can one measure the shape of a spontaneously symmetry-broken state of $^{238}$U or $^{20}$Ne? One cannot. The reason is simple: such a stationary state does not exist in nature and, therefore, is unavailable for experimentation. In quantum mechanical modeling, the deformed symmetry-broken state is a wave packet, a coherent superposition of states with different angular momenta. A possible experimental situation where such a state might briefly appear is in a heavy-ion fusion-evaporation reaction, where two ions fuse and rotate together at very high angular momentum. However, immediately after the first de-excitation photon is detected in a $\gamma$-ray detector, the wave packet collapses into a high-angular-momentum eigenstate, which then emits a series of $E2$ photons and rapidly relaxes to the $0^+$ ground state. In conclusion, the broken-symmetry states are not the ones examined in the UHIC experiments.

At this point, we are ready to discuss what is measured in the UHIC experiments. The key point is that one does not measure one-body nuclear moments in these events.
Instead, through event reconstruction, the positions of individual nucleons are identified based on measurements of outgoing particles. Multiple identical collisions then provide access to different quantum probability distributions. In short, every UHIC takes a snapshot of the spatial positions of the nucleons, not a snapshot of nuclear rotations or vibrations. In analogy, this can be related
to Coulomb explosion experiments, which
yield geometrical images of individual molecules \cite{Vager1989,Miller2022}.

Any measurement of the probability distribution to find one nucleon at any given position in space maps the (exact) one-body density of the nucleus. It is equal to the modulus squared of the (exact) many-body wave function, which depends on the coordinates of $A$ particles, integrated over $A-1$ coordinates. The one-body density of any 0$^+$ state is spherical. To determine the shape, one must use the concept of conditional probability~\cite {Romanovsky2006}, that is, the probability of finding one nucleon at a specific position in space under the condition that another nucleon occupies another position in space. Such a conditional probability is related to the (exact) two-body density (the integral of the exact wave function's modulus squared over $A-2$ coordinates). It has an axial shape because it must be symmetric to rotations about the line defining the position of the first point in question, and it can provide the density map in the plane spanned by the positions of both points. In this way, the axial shape of a 0$^+$ state can, in principle, be determined. Therefore, if the UHIC measurements can extract such conditional probabilities, the axial deformation can also be assessed in a 0$^+$ state. However, the one-body densities, as used in Refs.~\cite{Bally2022,Bally2023,Ryssens2023}, are insufficient for that purpose.

Examples of earlier theoretical analyses, involving two-particle correlations, can be found in, e.g., Refs.~\cite {Romanovsky2006,Sheikh21,Shen2023}, where the mean-field, symmetry-breaking, and symmetry-restored solutions are compared with the exact solutions of the problem, with many aspects of the arguments mentioned above explicitly presented. It is only recently that the nuclear UHIC modelers realized the importance of the many-body nuclear density for determining nucleons’ position distributions Refs.~\cite{(Gia23),(Dug25),(Bla25)}, and practical calculations of correlated samplings of nucleon positions were carried out in only one case study~\cite{Giacalone2025}.

The two-particle densities from UHIC imaging must be compared to those from modeled wave functions. However, this method for experimentally determining nuclear multipole moments is indirect because the values heavily depend on the model, and the impact of this dependence on uncertainty estimates remains unknown. As a result, they are not competitive with other methods that have been used so far, see Sec.~\ref{deformations}.

Additionally, the quantum mechanical spectroscopic quadrupole moment derived from the multipole expansion of the two-body density and the one derived from the one-body density are different observables. Notably, the first is measured in UHIC experiments, while the second is studied in low-energy experiments. Specifically, the first is accessible for $J<1$ states, whereas the second is not. Both may provide complementary but not necessarily identical information about nuclear ground states. This fact seems to have been overlooked in all existing experimental and theoretical studies of the problem. Theoretically, for $J\geq1$ states, both moments can be calculated and compared using symmetry-conserving {\it ab initio} or symmetry-restored mean-field methods. The analysis along these lines is in progress~\cite{(Sun25c)}.

In the context of the present discussion, it is helpful to note that a local approximation to the two-particle correlation function, based on the density-matrix expansion of the two-body density, is the nucleonic localization function \cite{Becke1990}. This function was initially introduced to visualize bond structures in molecules. The analogous nucleonic localization function was employed in mean-field calculations to study cluster configurations in nuclei \cite{Reinhard2011,Nakatsukasa2023}, cluster structures in fission \cite{Zhang2016}, and formation of clusters in both heavy-ion reactions \cite{Schuetrumpf2017} and high-spin states \cite{Li2020}.

In the same way, to calculate or measure triaxial deformations, one must consider the (exact) three-body density or measure correlations between positions of three nucleons~\cite{(Gia23)}. This differs from what was done in Ref.~\cite{(Jia22),Bally2022}, where nuclear triaxiality was inferred based on UHIC data.

In summary, ultra-relativistic collisions take place in the {\it laboratory frame}. Since $^{238}$U is spherical in its $J=0$ ground state and angular momentum is conserved, a deformed collision image is impossible, based on physics principles, regardless of the collision mechanism. (Likewise, a direct laboratory-system measurement of the electric dipole moment of the ammonia molecule in its ground state is impossible as the molecule has good parity~\cite{(And72)}.)

In other words, in the beam's coordinate system, the heavy ion {\em impinging on target} is described by a rotationally-invariant many-body wave function that depends on the coordinates of the nucleons, not just on the nucleon densities oriented in space.
It is {\em following the collision event} that the wave function collapses into one symmetry-breaking component, characterized by the nucleons' positions (planar image of the wave function in the coordinate representation).
Hence, to extract the initial QGP configuration, a simple one-body picture of the deformed nuclear density is insufficient.

\section{Static shapes and dynamic shape fluctuations \label{fluctuations}}

Publications by the STAR collaboration \cite{Abdulhamid2024,STAR2025} introduce the notion of ``instantaneous shapes'' and  ``long-timescale quantum fluctuations, making direct observation
challenging.'' It is claimed that ``Nuclear shapes, even in ground states, are not fixed. They exhibit
zero-point quantum fluctuations involving various collective
and nucleonic degrees of freedom at different timescales.
These fluctuations superimpose on each other in the laboratory
frame'' \cite{Abdulhamid2024}.

We wish to point out that there are no quantum fluctuations in any eigenstate of any quantum system. The shape of a projectile/target nucleus entering the collision does not fluctuate; there is no time-dependence involved, and, thus, there is no timescale. One should not confuse the model of a collective rotational or vibrational state, described as a time-dependent Slater determinant, with the reality of a quantum state that is accessible in an experiment. Similarly, one should not confuse dynamical fluctuations with the quantum uncertainty of an observable. An observable, the quantum quadrupole moment operator, defines the nuclear shape. The nuclear ground state is not an eigenstate of the quadrupole operator; therefore, any measurement of the quadrupole operator in the nuclear ground state must lead to the standard quantum dispersion of the results. Such dispersion is not related to any collective motion.

The effects describing the dispersion of static observables in the ground states of nuclei are well understood in nuclear structure. They are adequately accounted for by symmetry restoration after mixing the symmetry-broken states~\cite{(Rin80b),(Sch19c)}. Although they often go by the name of the ``vibrational'' corrections or ``fluctuations'' (quadrupole, octupole, pairing, etc.), they model the  stationary nuclear states. Such an approach to modeling UHIC results is in its infancy, see, e.g., ~\cite{Bally2022,Giacalone2025b, Li2025}. Still, it is a proper avenue to take, provided it is used to determine two-body densities and not only the multipole moments.

While a theoretical jargon exists pertaining to static deformations
(related to symmetry-breaking deformed mean-field solutions) and dynamical zero-point effects, or fluctuations (representing the dynamical beyond-mean-field corrections), the corresponding many-body wave functions are always stationary. Likewise, the jargon referring to the ``intrinsic'' reference frame is unhelpful in this context; it is better to use the proper quantum-mechanical notion of a symmetry-broken state.

\section{Nuclear deformations: a concept not entirely unknown\label{deformations}}

Nuclear deformations are parameters characterizing the anisotropy of the nuclear shape. These characteristics are not fundamental nuclear properties, as they
are deduced from various measured observables in a model-dependent way  \cite{NazRag}.
In low-energy nuclear physics, nuclear shapes have been characterized for decades using observables, such as quadrupole moments, obtained from a wide variety of experimental techniques. Since the seminal works by Sch{\"u}ler and Schmidt \cite{Schuler1935}  and Casimir \cite{Casimir1935,Casimir1936}, nuclear electric quadrupole moments $Q$ have been extracted, evaluated, and tabulated for nuclear ground states or excited states with total angular momenta of $J>1/2$, from a number of high-precision measurements \cite{Stone2021,Pyykko2018}, often quoted with uncertainties of 1\% or less for stable nuclei. The nuclear quadrupole moment $Q$ measures the deviation of the nuclear charge distribution from sphericity in a given nuclear state. In fact, Ref.~\cite{Stone2021} lists no less than 50 experimental techniques that form the basis for the $Q$ moments compiled. These range from the hyperfine splitting in mesic atoms \cite{Fitch1953,Cooper1953,Wheeler1953}, to a variety of laser spectroscopic techniques that are even applicable to short-lived rare isotopes \cite{Neyens2003, Yang2023}, to scattering approaches such as electron scattering \cite{Hofstadter1957,Donnely1975,Lightbody1972} or heavy-ion-induced Coulomb excitation \cite{Cline1986, Zielinska2022}. For a recent review on the history of nuclear shapes and experimental methods, the reader is referred to Ref.~\cite{Verney2025}.

This body of precision data from low-energy nuclear physics experiments seems to be overlooked by the UHIC community. Ref.~\cite{Bally2022}, for instance, questions the interpretation of spectroscopic experiments for odd-$A$ and odd-odd nuclei even though such systems have yielded precise $Q$ moments which ultimately quantify the deviation from sphericity.
One example from Ref.~\cite{Stone2021} is the $Q$ moment, $Q=+0.547(16)$b, quoted for the $3/2^+$ ground state of the odd-mass \nuc{197}{Au} nucleus from muonic x-ray spectroscopy, a nucleus that was much discussed by the UHIC community, e.g., in Refs.~\cite{Bally2023,Abdulhamid2024}. Additionally, taking any odd-odd nucleus, here \nuc{28}{Na}, from Ref.~\cite{Stone2021} reveals that $Q$ moments can be uniquely determined from spectroscopic experiments, e.g., for \nuc{28}{Na} from $\beta$-NMR spectroscopy where a high degree of nuclear polarization was achieved via optical pumping with a collinear laser beam \cite{Keim2000}. We note that this was accomplished for a short-lived rare isotope which cannot be accessed by UHIC at RHIC or the LHC.

For most stable and many short-lived nuclei, including even-even nuclei, which have a zero ground-state total angular momentum, and  odd-mass systems, a vast body of low-energy data exists from Coulomb excitation \cite{Cline1986, Zielinska2022} as well as from electron scattering \cite{Cooper1976,Phan1989}, sometimes even from M\"ossbauer spectroscopy \cite{Meeker1974}, on the deformation of excited states, including for quadrupole \cite{NNDC,Boris2016} and higher-order deformations such as the octupole (see Refs.~\cite{Butler1996,Butler2020}) and hexadecapole ones (e.g., \cite{Eichler1973,Reg1977a,Reg1977b,Wollersheim1977}). When appropriate in the scientific context, the results from such low-energy measurements use the $\beta_2$, $\beta_3$, $\beta_4$ and $\gamma$ \cite{Wu1996,Cline1986} parameterizations of the deformation as adopted by the UHIC community. Often, the direct use of reduced electromagnetic transitions strengths, $B(E2)$, $B(E3)$, and $B(E4)$ proved to be more insightful and scientifically appropriate in the context of the field. The case has yet to be made for a unique nuclear structure insight from UHIC on \nuc{238}{U} \cite{Abdulhamid2024,Ryssens2023,Xu2024}, \nuc{20}{Ne} \cite{Giacalone2025,Giacalone2025b}, \nuc{96}{Zr} and \nuc{96}{Ru} \cite{Zhang2022}, \nuc{129}{Xe} \cite{Bally2022}, \nuc{197}{Au} \cite{Bally2023}, and \nuc{150}{Nd} \cite{Li2025}, given the substantial and consistently evaluated body of low-energy data on nuclear deformation \cite{NNDC,Stone2021,Boris2016}.

Theoretically, nuclear deformations can be obtained from calculated one-body nuclear densities of symmetry-broken states, including charge density, proton density, or neutron density.
A substantial body of literature exists on local and global predictions of nuclear shape deformations, primarily obtained with various flavors of mean-field theory and its extensions. Global surveys can be found in, e.g.,  Refs.~\cite{Moller2008,Moller2016,Goriely2016,Erler2012,Agbemava2014,Cao2020,Zhang2022a,Guo2024,Zhou2025} and the theoretical databases MassExplorer \cite{massexplorer2} and BrusLib \cite{Bruslib}. As discussed in Ref.~\cite{NazRag}, there is a weak model dependence of nuclear deformations of well-deformed nuclei as they reflect the geometries of the valence Hartree-Fock (Kohn-Sham, canonical, Nilsson) orbits.

The notion of nuclear deformation is also not robust: for many nuclei in the nuclear landscape, such as transitional \cite{Faessler1982} or shape-coexisting \cite{Wood1992,Heyde2011} systems, where there are fundamental problems with defining shape deformations \cite{Reinhard1984,Nazarewicz1994}.

\section{Uncertainty quantification}\label{UQ}

Uncertainty quantification is at the heart of any pipeline connecting models that span disparate scales.
By quantifying the models' discrepancies and deficiencies, one can construct a rigorous statistical framework that provides a means to transfer information both forwards and backwards within this multiscale (or multifidelity) modeling framework.
In the context of the current work, the forward direction involves propagating uncertainties from quantified nuclear models through to final-state observables in UHIC.
The natural next step is then to have information flowing in the other direction in the pipeline.
This provides an opportunity to solve the inverse problem and perform statistical inference on shapes and other correlated observables accessible within the chosen nuclear structure model.
Both steps represent exciting developments and are  examples of the confluence of advances in theoretical modeling, experimental analyses, and computational statistics and data science in modern nuclear physics.

A caveat emerges, however, when one attempts to enact this flow of information for complex multiscale models.
There has been considerable progress in quantifying the uncertainty in the context of UHIC in recent years: see Refs.~\cite{Nijs2021,Ehlers2024,STAR2025} for a non-exhaustive selection of works with a detailed discussion of UQ practices.
However, without careful consideration of the full uncertainties for each component of the entire multistage process, one runs the risk of misspecified uncertainties polluting parts of the pipeline.
Fits and calibrations, in general, will lead to distributions of tunable parameters that allow one, within a given credible interval, to explain an equivalent subset of available data.
If those tuneable parameters control a model with an intrinsic model deficiency or inability to describe different classes of observables simultaneously, the interpretation of those model outputs and parameters must be viewed with skepticism.

This fact is particularly concerning if one wishes to use information from UHIC alone to constrain nuclear structure models or to provide information on nuclear shapes.
In the simplest applications of this framework, the nuclear structure model has been chosen as a deformed Fermi form factor with deformation parameters constraining the bulk shape of the nucleus.
In this example, the calibration of the model involves exploring the various parameters that define the shape and determining a statistical ensemble of shapes that best agree with the downstream UHIC data.
Without a simultaneous constraint or consideration of binding energies, charge radii, or deformations from other observables, it is difficult to know the full extent of the discovery potential of such analyses.

It is also essential to acknowledge the fact that any conclusion drawn in such an analysis is inherently model-dependent.
The experimental data, on an event-by-event basis, undoubtedly encode intricate details about the correlated nuclear system.
Decoding this information, however, necessitates a sophisticated theoretical modeling framework that translates UHIC observations
into quantitative measures of nuclear deformation.
This framework typically involves several interconnected stages, each with its own set of assumptions and uncertainties.

Centrality, for instance, in heavy-ion collisions is typically determined by measuring the number of charged particles produced in the collision and then comparing this number to a Monte Carlo Glauber calculation or another similar framework~\cite{ALICECentrality2013,Loizides:2014vua,ALICE:2018tvk,Aaij_2022}.
This approach, while widely used, has certain limitations.
The accuracy of centrality determination relies on the validity of the chosen model, which has inherent assumptions about the nucleon-nucleon interactions and the density distribution of nucleons within the nucleus.
Additionally, the efficiency of detecting charged particles and the methods used to distinguish between particles originating from the collision and those from secondary decays can introduce uncertainties.
The subsequent hydrodynamic evolution of the QGP is another example, as it relies on equations of state and transport coefficients that are not fully constrained by experimental data, and even then, there are multiple competing models to propagate the QGP after formation~\cite{shen2020recent}.
In addition to this hydrodynamical flow, there are also nonflow correlations that must be corrected for and that add further model-based uncertainties to the analysis pipeline~\cite{Feng_2025}.
The recent debate~\cite {wang2024nonflowissueconnectinganisotropy,STAR2024} underscores the importance of these nonflow contributions, specifically in the context of nuclear shape extraction.
Finally, the process of hadronization, whereby the QGP converts back into hadrons, introduces further uncertainties related to the fragmentation, clusterization, and subsequent decay of these particles~\cite{ONO2019139,altmann2025towards}.

The complexity and non-uniqueness of this modeling framework raise concerns regarding the robustness and reproducibility of results drawn from the method.
Different research groups, employing subtly different implementations of the same underlying stages, could potentially arrive at different conclusions regarding the extracted deformations of the colliding system.
These discrepancies can arise from variations in the specific data sets considered in the fit, the choice of initial conditions, the details of the hydrodynamic evolution, the treatment of hadronization, and other factors.
While this model dependence has not been investigated in detail for shape extraction, model discrepancy studies have been performed for UHIC-related quantities in Ref.~\cite{Jaiswal2025}, and the impact on the Bayesian inference was found to be significant.

Finally, it is worth discussing the required properties and fidelity of the nuclear structure models to be useful in this pipeline.
At a minimum, one should consider `reasonable' models of nuclei.
A convenient definition in this case would be any model that allows one to simultaneously describe (and predict)  observables in low-energy nuclear structure to a reasonable accuracy.
For the forward direction in the pipeline, this would also ideally involve quantified models that have been calibrated to a wide class of nuclear observables and have been validated on the class of observables that are of most interest in any particular study.
The uncertainties of the initial state should then be propagated
through the pipeline, from the hydro phase of the QGP to the hadronization phase.
This already provides an excellent anchor point for UHIC studies to ensure that the nuclear structure input is well validated, providing further confidence in any UQ efforts for the UHIC simulation pipeline itself.

For the reverse direction, the core question is ``are the constraints and resulting uncertainties sufficient to distinguish between models used to predict the initial state?''.
If the discussed pipeline is rather coarsely constraining, then one should only expect differences to arise from large-scale, bulk changes in the nucleus.
For models meeting the reasonableness criteria above, these bulk properties are likely to be very similar.
Furthermore, employing any nuclear structure models for inference on nuclear properties of any sort {\em requires} quantified Hamiltonians and detailed model discrepancy studies to validate those models on nuclear deformations and to render any conclusions statistically significant.
This is equally true for any model of nuclear structure, including {\it ab initio} models.
While this discussion has focused on nuclear shapes, this is a core point for the extraction of any properties of nuclei.
For example, the investigation in Ref.~\cite{Li2025} suggests UHIC experiments as a novel probe of matrix elements central to the search for neutrinoless double-beta decay, although without considering the constraining potential of those UHIC observables in the context of other available nuclear structure observables (see, e.g., Ref .~\cite{Cirigliano2022} for a multitude of factors
impacting matrix elements for neutrinoless double-beta decay).
More fundamentally, one needs to demonstrate the suitability of the structure model itself for such studies by showing that, when considering data from other sources, the results are not in conflict with well-known nuclear properties.

\section{Conclusions \label{Conclusions}}

``Extraordinary claims require extraordinary evidence'' \cite{Sagan1979}.
The purpose of this paper is to shed light on the collective-flow-assisted nuclear shape imaging employed in the analysis of data from UHIC, and to assess the claims of high relevance of these experiments to nuclear structure research.
While there is little doubt that the proper treatment of the  geometries of colliding heavy ions is essential for the characterization of the initial QGP configuration
and, hence, the extraction of fundamental properties of QGP from UHIC, we believe that the proposed nuclear shape-imaging method is based on several flawed assumptions and, hence, its usefulness as a ``discovery tool for
exploring nuclear structure'' \cite{Abdulhamid2024}  has been overstated.

{\em First}, the modeling of the initial QGP state formed following the heavy-ion collision involves the description of the relative position identification of the individual nucleons. To this end, the information contained in the one-body nucleonic density, as presented in the majority of papers dealing with nuclear imaging from UHIC, is insufficient. The proper tool for this imaging of the nucleon's position distributions is the two-body density (to map axial distributions) or the three-body one (to map triaxial shapes).
The schematics depicting the deformed orientations of nuclei are misleading, as even-even nuclei, such as $^{238}$U or $^{20}$Ne, have isotropic ground-state wave functions in the laboratory system. In this context, the recently proposed approaches in
Refs.~\cite{(Gia23),(Dug25),(Bla25),Giacalone2025,Ke2025}, based on the many-body wave function, hold promise.

{\em Second}, the notions of ``instantaneous shapes,'' ``long-timescale quantum fluctuations,'' and ``zero-point quantum fluctuations involving various collective
and nucleonic degrees of freedom at different timescales'' \cite{Abdulhamid2024} make little physical sense as the many-body wave function of the colliding ion is stationary.

{\em Third}, rich databases exist of shape deformations measured in low-energy experiments, as well as of theoretical deformations computed within nuclear mean-field theory. The data on stable nuclei are particularly rich and seem more than sufficient to inform models of the initial QGP state.

{\em Fourth,} the existing precise information on nuclear shapes should be utilized by nuclear UHIC imaging practitioners to validate analysis techniques and control model uncertainties and discrepancies. Currently, a comprehensive and combined uncertainty quantification analysis of UHIC data has not been carried out because of the lack of a consistent methodology
for propagating uncertainties from  nuclear models through to final-state observables, and with propagating uncertainties
backwards in the pipeline.

In summary, we find that the current use of UHIC to image low-energy nuclear structure – namely, nuclear deformations – is prone to many flaws of interpretation and precision. Nuclear one-body electric moments for stable or long-lived nuclei are well understood and have been studied extensively over many years using a wide range of techniques, both direct and indirect. This is still a very active area of research, as many detailed challenges remain in low-energy experiment and theory. The UHIC measurements do not directly contribute to this field of study because they provide many-body correlations in the laboratory frame. As proposed in this work, there is potential for using the measured many-body correlations by the UHIC nuclear imaging to examine multipole collectivity from an entirely new perspective. This could open up a fascinating new research avenue.

The modeling dependence, large uncertainties, and limited isotopic reach of UHIC-based studies pose significant challenges and hinder the method’s ability to meaningfully add to the existing corpus of data on nuclear shapes. On the other hand, the existing systematic nuclear structure measurements and calculations provide strong constraints for the modeling of the initial state of UHIC and a firm baseline for future studies of QGP formation and flow. With that being said, a compelling feature of UHIC analyses is that they explicitly probe many-body correlations in the nucleus. Focusing on this aspect, rather than shapes, represents a strength of the approach worth playing to.

\bigskip
\begin{acknowledgments}
This material is based upon work supported by the U.S. Department of Energy under Award Numbers DE-SC0013365 and DE-SC0023633 (Office of Science, Office of Nuclear Physics), DE-SC0023175 (Office of Science, NUCLEI SciDAC-5 collaboration), DE-FG02-97ER41041, DE-FG02-97ER41033, and the National Science Foundation under award number 2004601 (CSSI program, BAND collaboration).
This work was also partially supported by the STFC Grant Nos.~ST/V001035/1 and~ST/Y000285/1.

\end{acknowledgments}


%

\end{document}